\documentclass[rmp,amsmath,amssymb]{revtex4} 
\usepackage{graphicx} 
\usepackage{epsfig} 
\usepackage{ae}
\usepackage{aecompl}
\usepackage{color}
\usepackage[normalem]{ulem}

\makeatletter
\def\bgcolour{\@ifnextchar[\bgcolour@i{\bgcolour@ii{}}}
\def\bgcolour@i[#1]{\bgcolour@ii{[#1]}}
\def\bgcolour@ii#1#2{%
  \bgroup%
  \markoverwith{%
    \hbox{%
      \begingroup%
      \kern-.05em%
      \color#1{#2}%
      \strut%
      \vrule\@width.3em%
      \kern-.05em%
      \endgroup%
    }%
  } \ULon%
}
\makeatother

\def\highlight{\bgcolour{white}}

\newcommand{\bfr}{{\bf r}}
\newcommand{\bw}{{\bar w}}
\newcommand{\bphi}{{\bar\phi}}
\newcommand{\bxi}{{\bar\xi}}
\newcommand{\fwidth}{3.5in}

\begin{document}

\title{Field theoretic study of bilayer membrane fusion:\\
       I. Hemifusion mechanism.}

\author{K.\ Katsov$^1$, M.\ M\"{u}ller$^2$, M.\ Schick$^1$}
\affiliation{$^1$Department of Physics, University of Washington,  Box
  351560, Seattle, WA 98195-1560} 
\affiliation{$^2$Institut f{\"u}r Physik, WA 331, Johannes Gutenberg
 Universit{\"a}t, D-55099 Mainz, Germany}

\date{\today}

\begin{abstract} 

Self-consistent field theory is used to determine structural and energetic 
properties of metastable intermediates and unstable transition states 
involved in the standard stalk mechanism of bilayer membrane fusion. A 
microscopic model of flexible amphiphilic chains dissolved in hydrophilic 
solvent is employed to describe these self-assembled structures. We find 
that the barrier to formation of the initial stalk is much smaller than 
previously estimated by phenomenological theories. Therefore its creation 
it is not the rate limiting process. The barrier which is relevant is 
associated with the rather limited radial expansion of the stalk into a 
hemifusion diaphragm. It is strongly affected by the architecture of the 
amphiphile, decreasing as the effective spontaneous curvature of the 
amphiphile is made more negative. It is also reduced when the tension is 
increased. At high tension the fusion pore, created when a hole forms in 
the hemifusion diaphragm, expands without bound.  At very low membrane 
tension, small fusion pores can be trapped in a flickering metastable 
state.  Successful fusion is severely limited by the architecture of the 
lipids.  If the effective spontaneous curvature is not sufficiently 
negative, fusion does not occur because metastable stalks, whose existence 
is a seemingly necessary prerequisite, do not form at all. However if the 
spontaneous curvature is too negative, stalks are so stable that fusion 
does not occur because the system is unstable either to a phase of stable 
radial stalks, or to an inverted-hexagonal phase induced by stable linear 
stalks. Our results on the architecture and tension needed for successful 
fusion are summarized in a phase diagram.

\end{abstract} \maketitle	

\section{Introduction} \label{sec:intro} The importance of membrane fusion 
in biological systems hardly needs to be emphasized. It plays a central 
role in trafficking within the cell, in the transport of materials out of 
the cell, as in synaptic vesicles, and in the release of endosome-enclosed 
external material into the cell, as in viral infection. While proteins 
carry out many functions leading up to fusion, such as ensuring that a 
particular vesicle arrives at a particular location, or bringing membranes 
to be fused into close proximity, there is much evidence that they do not 
determine the actual fusion mechanism itself.  Rather the lipids 
themselves are responsible for the evolution of the fusion process in 
which the lipid bilayers undergo topological change 
\cite{Lee98,Zimmerberg99,Lentz00}.

The physical description of the fusion process has, until very recently, 
been carried out using phenomenological theories which describe the 
membrane in terms of its elastic moduli \cite{Safran94}.  The application 
of these theories to fusion has been reviewed recently 
\cite{Zimmerberg99}. The fusion path which has been considered by these 
methods is one in which local fluctuations are assumed to cause a 
rearrangement of lipids in the opposed {\em cis} leaflets, resulting in 
the formation of a stalk \cite{Markin83}. To release tension imposed on 
the membranes by the reduction of the solvent between them, the inner {\em 
cis} layers recede, decreasing their area, and bringing the outer, {\em 
trans}, leaves, into contact. In this way the stalk expands radially to 
form a hemifusion diaphragm. Creation of a hole in this diaphragm 
completes formation of the fusion pore.

In recent years, coarse grained models of amphiphiles \cite{Mueller003} 
have been used to provide a microscopic, as opposed to phenomenological, 
description of membranes. Fusion was studied within two such models.  
One, in which non-flexible molecules were composed of three segments, was 
studied by Brownian dynamics simulation \cite{Noguchi01}, and the other, 
in which the amphiphiles were modeled as flexible, polymer, chains in 
solvent, was studied by Monte Carlo simulation \cite{Mueller02}. Both 
models showed a markedly different path to fusion than the 
phenomenological approaches assumed. Along this new path, the creation of 
the stalk is followed by its non-axially-symmetric growth, {\em i.e.} 
elongation.  After the stalk appears, there is a great increase in the 
rate of creation of holes in either bilayer, and the holes are created 
near the stalk itself. After a hole forms in one bilayer, the stalk 
elongates further and surrounds it, forming a hemifusion diaphragm. 
Formation of a second hole in this diaphragm completes the fusion pore. 
Alternatively, the second hole in the other membrane can appear before the 
elongated stalk surrounds the first hole. In this case, the stalk aligns 
the two holes and surrounds them both forming the fusion pore.  This 
mechanism has also been observed more recently in molecular dynamics 
simulations \cite{Marrink03,Stevens03}. As has been stressed by us 
recently \cite{Mueller03}, this alternative mechanism can be distinguished 
experimentally from the earlier hemifusion mechanism because it predicts 
at least two phenomena that are not compatible with the earlier 
hypothesis. The first is lipid mixing between the {\em cis} leaf of one 
bilayer with the {\em trans} layer of the other, a phenomena which has 
been observed \cite{Evans02,Lentz97}. The second is transient leakage 
through the holes, noted above, which is correlated in space and time with 
the fusion process. Just such correlated leakage has recently been 
observed and extensively studied \cite{Frolov03}.

The Monte Carlo (MC) simulations of the fusion process showed very clearly 
the nature of the process, and obtained quantitative correlations between 
leakage and fusion. Unfortunately, simulations are computationally 
expensive, so that investigation of the fusion process for different 
molecular architectures and membrane tensions is impractical. Moreover 
simulations are not well suited for calculating free energy barriers of 
the fusion intermediates in the mechanism observed over a wide range of 
tension and amphiphilic architecture. Not only would one like to obtain 
these barriers, one would also like to compare them to those of the 
intermediates involved in the original hemifusion mechanism. To do so, we 
employ a standard model of amphiphilic polymers, which we describe in the 
next section, and solve it within the framework of Self-Consistent Field 
Theory (SCFT).  In the first paper of this series we examine the original 
hemifusion mechanism, while in the second we shall consider the new 
mechanism observed in the simulations.

In Sec.~\ref{sec:iso-bilayer} we present the basic properties of the 
isolated bilayers and monolayers which result from our calculation. These 
properties include the spatial distribution of hydrophilic/hydrophobic 
segments, the area compressibility, the bending rigidity and spontaneous 
curvature. We compare them to those obtained independently from previous 
Monte Carlo simulations, and from experiments on liposomes and 
polymersomes. It should be noted, that these effective {\em macroscopic} 
properties are calculated within our {\em microscopic} approach, and are 
not required as input, in contrast to the common phenomenological 
descriptions of fusion based on membrane elasticity theory.

\begin{figure}[!ht]
\includegraphics[width=\fwidth,height=7.25in]{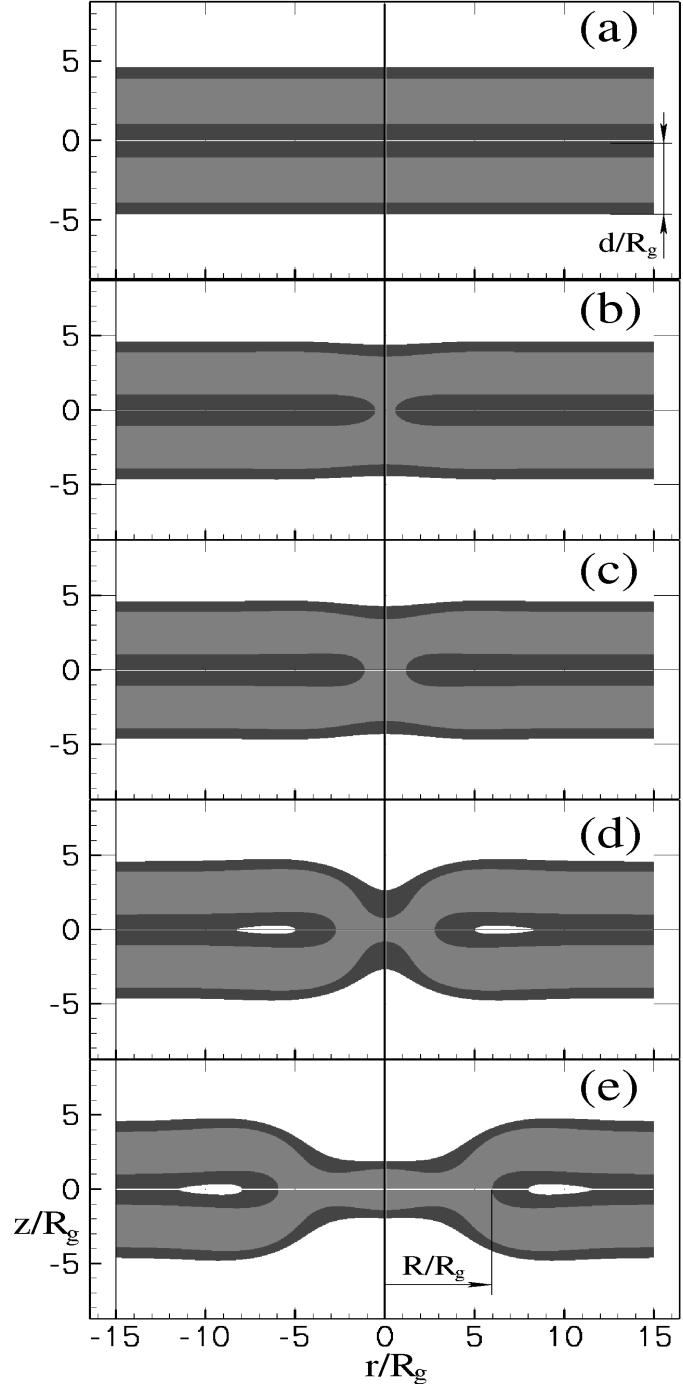}
\caption{Density profiles of the stalk-like structures shown in the $r,z$ 
plane of cylindrical coordinates. As the structures are axially symmetric, 
the figure and its reflection about the $z$ axis are shown for the 
viewer's convenience. The amphiphiles contain a fraction $f=0.35$ of the 
hydrophilic component. The bilayers are under zero tension.  Only the 
majority component is shown at each point: solvent segments are white, 
hydrophilic and hydrophobic segments of the amphiphile are dark and light 
correspondingly. 
\protect\highlight{Distances are measured in units of the polymer radius of 
gyration, $R_g$, which is the same for both the amphiphiles and 
for the homopolymer solvent.} (a) Two bilayers in solvent. There is no 
stalk between them. 
\protect\highlight{Their thickness, $d$, defined in the text, Eq. 
(\ref{thickness}) is shown.} (b) Unstable transition state to the formation 
of the initial stalk, a state we label $S_0$. The radius of the stalk is 
$R=0.6R_g$. In general the stalk radius $R$ is defined by the condition on 
the local volume fractions $\phi_A(R,0)-\phi_B(R,0)=0$. (c) The metastable 
stalk itself, which we label $S_1$. Its radius is $R=1.2R_g$. (d) The 
unstable transition state between the metastable stalk and the hemifusion 
diaphragm. This transition state is denoted $S_2$. Its radius is 
$R=2.8R_g$. (e) A small hemifusion diaphragm of radius $R=3.4R_g$. The 
radius is shown explicitly. } \label{prof-stalk} \end{figure}

\begin{figure}[!ht]
\includegraphics[width=\fwidth]{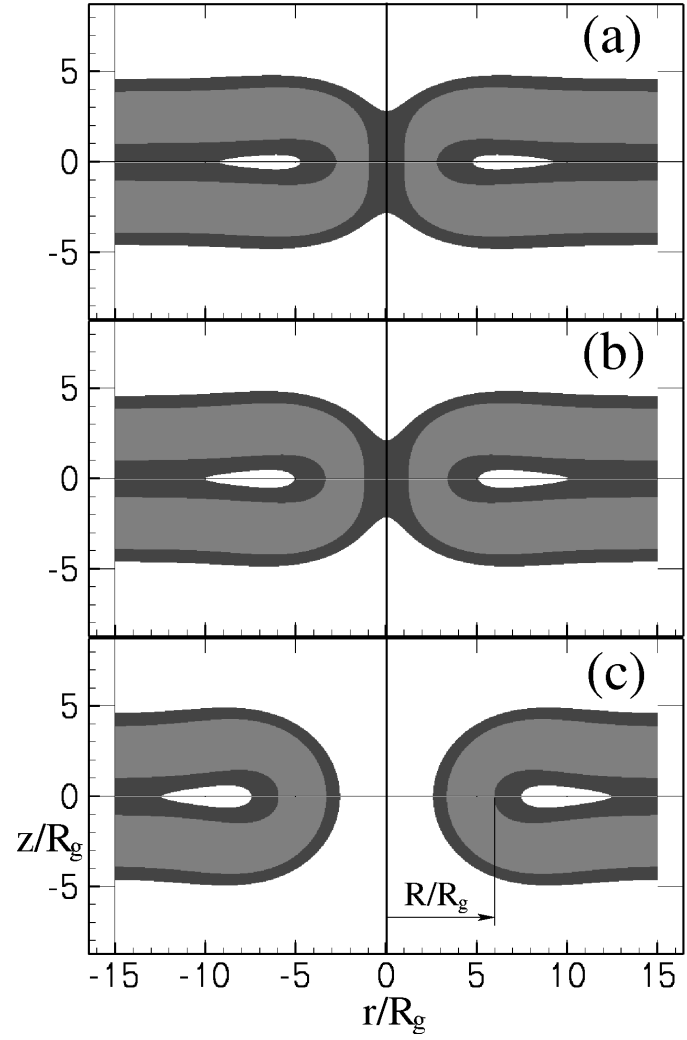} 
\caption{(a) Density 
profile of a fusion pore of radius $R/R_g=2.4$. Were the radius any 
smaller, the pore would be absolutely unstable to a stalk-like structure. 
In general the pore radius is the larger value of the radial coordinate 
which satisfies the condition on the local volume fractions 
$\phi_A(R,0)-\phi_B(R,0)=0$. (b) A pore of radius $R/R_g=3.4$ and (c) one 
of radius $R/R_g=6.0$. The radius is shown explicitly. The hydrophilic 
fraction of the amphiphile, $f=0.35$ as in Fig.1 and the tension is zero, 
again as in Fig. 1. Distances are measured in units of the radius of 
gyration $R_g$.} \label{prof-pore} \end{figure}

In Sec.~\ref{sec:stalk-mech} we examine the free energy landscape along 
the standard hemifusion pathway, one which is shown in 
Figs.~\ref{prof-stalk} and \ref{prof-pore}.  This is the same path as 
assumed in phenomenological approaches, but we use a microscopic, 
molecular, model to calculate along this trajectory the distribution of 
microscopic components in the system, and the free energy which results. 
The initial configuration of the system is that of two parallel bilayers 
(Fig.~\ref{prof-stalk}(a). Hydrophilic portions of the amphiphile are 
shown dark, hydrophobic portions are shown light, and solvent is shown 
white.  To bring the bilayers into close contact requires energy to reduce 
the amount of solvent between them. Consequently the free energy per unit 
area, or tension, of the bilayers increases.  Fusion is one possible 
response of the system to this increased tension. Along the standard 
pathway, close contact of the {\em cis} layers is followed by the creation 
of an axially-symmetric {\em unstable} intermediate, or transition state, 
shown in cross-section in 1(b), which leads, under conditions detailed 
below, to a {\em metastable}, axially-symmetric stalk, 1(c). If the system 
is under sufficient tension, this stalk pinches down, passes through 
another unstable intermediate, 1(d), and expands to form a hemifusion 
diaphragm 1(e).  In order to determine the radius at which a hole forms in 
this diaphragm to complete the fusion pore, we calculate in 
Sec.~\ref{sec:pore} the free energy of an axially-symmetric fusion pore as 
a function of {\em its} radius. Density profiles of fusion pores are shown 
in Fig.~\ref{prof-pore}. We assume that when the free energy of the 
expanding hemifusion diaphragm exceeds that of a fusion pore of the same 
radius, a hole forms in the diaphragm converting it into a fusion pore, 
Fig.~\ref{prof-pore}(b). Not surprisingly, the radius of the fusion pore 
which is formed can not be too small, as in Fig ~\ref{prof-pore}(a), for 
then its free energy would be higher than that of the hemifusion 
diaphragm. Once the pore has formed, it will expand if the membranes are 
under tension. Fig.~ \ref{prof-pore}(c)  shows the profile of an expanded 
pore, although this one is in a membrane of zero tension. We are able to 
calculate the free energy of the system at all stages of the pathway for 
various architectures of the amphiphile and tensions of the membrane.

The most notable results of our calculation are as follows. 
\begin{list}{}
\item i. The free energy barrier to form a stalk, that is, the free energy
difference between the initial configuration of Fig.~\ref{prof-stalk}(a)  
and that of the \highlight{transition state}
 Fig.~\ref{prof-stalk}(b), is small,
on the order of 10 $k_BT$, much smaller than the estimates of
phenomenological theories \cite{Kuzmin01}.
\item ii. The more important fusion barrier is encountered on the path
between the metastable stalk, Fig.~\ref{prof-stalk}(c), through 
\highlight{another transition state}, 
Fig.~\ref{prof-stalk} (d), to a small hemifusion
diaphragm, Fig.~\ref{prof-stalk}(e). The height of this barrier depends
strongly on both the effective spontaneous curvature of the amphiphile, as
well as on the membrane tension. As expected, the barriers to fusion are
reduced as the architecture is changed so as to approach the transition
from the lamellar to the inverted-hexagonal phases. The effect of tension
on the barriers to fusion is less dramatic, but still very important.
\item iii. We find that the hemifusion diaphragm does not expand
appreciably before converting to the fusion pore.
\item iv. The small fusion pore formed by rupturing the hemifusion
diaphragm can, at low tension, be trapped in a metastable state and not
expand further. This result provides an explanation for the flickering
fusion events which are observed experimentally. As the tension is
increased, however, this metastable state disappears, and the fusion pore,
once formed, expands without limit, thus resulting in complete fusion.
\item v. We observe that {\em the regime of successful fusion is rather
severely limited by the architecture of the lipids.} If their effective
spontaneous curvature is too negative, fusion is pre-empted by the
formation of either an inverted-hexagonal phase or of a stalk phase. If
their curvature is not sufficiently negative, fusion is prevented by the
absence of a metastable stalk.  
\end{list}

We discuss these results further in
Sec.~\ref{sec:disc}.

\section{The model}
\label{sec:model}

We consider a system consisting of an incompressible mixture of two kinds 
of polymeric species contained in a volume $V$.  There are $n_a$ 
amphiphilic diblock copolymers, composed of $A$ (hydrophilic) and $B$ 
(hydrophobic) monomers, while the $n_s$ solvent molecules are represented 
by hydrophilic homopolymers consisting of $A$ segments only. The fraction 
of hydrophilic monomers in the diblock is denoted $f$, and the identical 
polymerization indices of both the copolymer and the homopolymer are 
denoted $N$.  The hydrophilic and hydrophobic monomers interact with a 
local repulsion of strength $\chi$, the Flory-Huggins parameter. Provided 
that these parameters are chosen appropriately, this model is essentially 
equivalent to that which we simulated earlier \cite{Mueller03}. We 
emphasize that the earlier simulation and the calculation presented here 
are {\em completely independent.} Comparison of results obtained from the 
two different calculations, therefore, are instructive.

The partition function of the system of flexible chains with Gaussian 
chain statistics can be formulated easily in either canonical or grand 
canonical ensembles \cite{Matsen95,Schmid98}, but is too difficult to be 
evaluated analytically.  Consequently we employ the well established 
Self-Consistent Field Theory, (SCFT), to obtain a very good approximation 
to the partition function. This theory has been recently reviewed \cite 
{Schmid98}, so we relegate to the Appendix a brief reminder of the salient 
features of the approximation. It suffices here to say that the 
inhomogeneous systems we shall be studying are characterized by a local 
volume fraction of hydrophilic units, $\phi_A$, and of hydrophobic units, 
$\phi_B$. Due to the incompressibility constraint, these two volume 
fractions add to unity locally. It is convenient to control the relative 
amounts of amphiphile and of solvent, and therefore the relative amounts 
of $A$ and $B$ monomers, by an excess chemical potential $\Delta\mu$, the 
difference of the chemical potentials of amphiphile and of solvent.  
Within this ensemble, the Self-Consistent Field Theory produces a free 
energy, $\Omega$, which is a function of the temperature, $T$, the excess 
chemical potential, $\Delta \mu$, the volume $V$, and, when bilayers are 
present, the area $A$ which they span. In addition, the free energy is a 
{\em functional} of the local volume fractions $\phi_A$ and $\phi_B$ which 
are obtained as solutions of a set of five non-linear, coupled equations, 
given in the Appendix, Eqs. \ref{wa} to \ref{SCFT-eqn}.  Acceptable 
solutions are defined by various constraints. For example, we require that 
all solutions be axially symmetric, an assumption embedded in all previous 
treatments of the stalk/hemifusion process. We take the axis of symmetry 
to be the $z$ axis of the standard cylindrical coordinate system, 
$(r,\varphi,z)$. Further, we require all solutions to be invariant under 
reflection in the $z=0$ plane.  Thus 
$\phi_A(r,\varphi,z)\rightarrow\phi_A(r,z)=\phi_A(r,-z),$ and similarly 
for $\phi_B$. Other constraints are more interesting. For example, to 
describe a stalk-like structure, as in Fig.~\ref{prof-stalk}, we require 
that the solution display a connection, along a portion of the $z$ axis, 
between the hydrophobic portions of the two bilayers.  Further, we 
constrain its radius, $R$, defined by the condition 
$\phi_A(R,0)=\phi_B(R,0)$, to be a value specified by us. In this way we 
are able to calculate the free energy of stalk-like structures as a 
function of their radius.  To reiterate, once a solution of the set of 
non-linear coupled equations is found which, for given temperature, excess 
chemical potential, and volume, satisfies the various constraints imposed, 
the free energy within the Self-Consistent Field Theory follows directly. 
We now turn to a description of the results of this procedure for the 
systems of interest. We begin with those obtained for isolated bilayers.

\section{Properties of Isolated Bilayers} \label{sec:iso-bilayer} 
\begin{figure}[!ht]
\includegraphics[width=\fwidth]{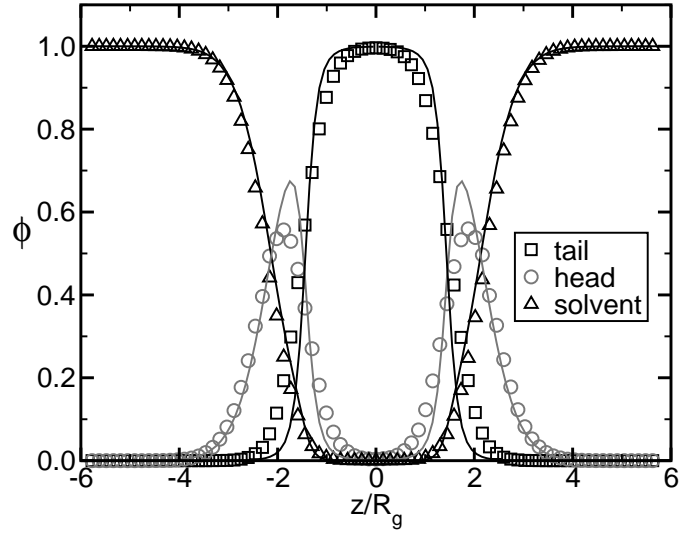}
\caption{Results from SCFT (solid lines) for the composition profile 
across a bilayer membrane which is under zero tension. For comparison, the 
profile obtained independently from simulation by averaging over 
configurations is shown in symbols. The hydrophilic fraction of the 
amphiphiles is $f=0.34$ } \label{comp-vs-z-LT} \end{figure}

In this 
section, we present a range of microscopic and thermodynamic properties of 
isolated bilayer membranes 
\highlight{in excess solvent} 
which follow from our 
calculation. The most basic structural properties of the bilayer membrane 
is the distribution of the hydrophilic and hydrophobic segments across it.  
In Fig.~\ref{comp-vs-z-LT} we present the composition profiles obtained 
within the SCFT approach, shown by solid lines, and compare them to those 
obtained from our independent simulations \cite{Mueller03} which are shown 
by the symbols. The profiles from simulation are averaged over all 
configurations. In the Monte Carlo simulation we used amphiphilic polymers 
consisting of $11$ hydrophilic and $21$ hydrophobic monomers. This 
corresponds to a hydrophilic fraction $f=11/32\approx 0.34$, the value we 
also used in the SCFT calculations. Fig.~\ref{comp-vs-z-LT}, corresponds 
to a system under zero tension. The profiles change quantitatively, but 
not qualitatively, as the tension is increased.

The overall agreement between the SCFT and averaged MC simulation results 
is very good. The position and the width of the regions enriched in $A$ 
(head) and $B$ (tail) segments of the amphiphile and solvent segments are 
reproduced quantitatively in the SCFT model.  The small discrepancies in 
the $A/B$ interfacial width can be attributed to capillary waves and 
peristaltic fluctuations present in the MC simulations, but neglected in 
the SCFT calculations.

Thermodynamic properties of the bilayer can be calculated from the free 
energy of the system which contains such a bilayer of area ${\cal A}$. We 
denote this free energy $\Omega_m(T,\Delta\mu,V,{\cal A })$.  Similarly, 
we denote the free energy of the system without the membrane, i.e., a 
homogeneous amphiphile solution, $\Omega_0(T,\Delta\mu,V)$.  The 
difference between these two free energies, in the thermodynamic limit of 
infinite volume, defines the excess free energy of the membrane: 
\begin{equation} \delta\Omega_m(T,\Delta\mu,{\cal A})\equiv 
\lim_{V\rightarrow\infty}[\Omega_m(T,\Delta\mu,V,{\cal A})- 
\Omega_0(T,\Delta\mu,V)]. \end{equation}

The excess free energy per unit area, in the thermodynamic limit of
infinite area, defines the lateral membrane tension
\begin{equation}
\gamma(T,\Delta\mu)\equiv\lim_{A\rightarrow\infty} 
[\delta \Omega_m(T,\Delta\mu,A)/ A].
\end{equation}
In the grand canonical ensemble this tension $\gamma$
can be related to the temperature and chemical potential by means of the
Gibbs-Duhem equation  

\begin{equation} {\rm d}\gamma(T,\Delta\mu)=-\delta s\ {\rm d}T-\delta
\sigma_a {\rm d}(\Delta\mu), \end{equation}
where $\delta s$ is the excess entropy per unit area, and $\delta\sigma_a$
is the excess number of amphiphilic molecules per unit area. This relation
is quite useful because it shows that one can set the tension to any given
value by adjusting the excess chemical potential of amphiphiles at
constant temperature.

Depending on the applied tension, the thermodynamic behavior of the 
membrane can be classified into three generic types. For values of the 
tension which are sufficiently small and positive, the membrane is 
metastable with respect to rupture, and it is this range of tension which 
we consider below. For larger positive values of $\gamma$, the membrane 
becomes absolutely unstable to rupture, while for negative $\gamma$, the 
system is unstable to an unlimited increase in the membrane's area, which 
simply leads to formation of the bulk lamellar phase.  In the following, 
we shall use the dimensionless tension, $\gamma/\gamma_{\rm int}$, where 
$\gamma_{\rm int}$ is the interfacial free energy per unit area between 
coexisting solutions of hydrophobic and hydrophilic homopolymers at the 
same temperature.

\begin{figure}[!ht]
\includegraphics[width=\fwidth]{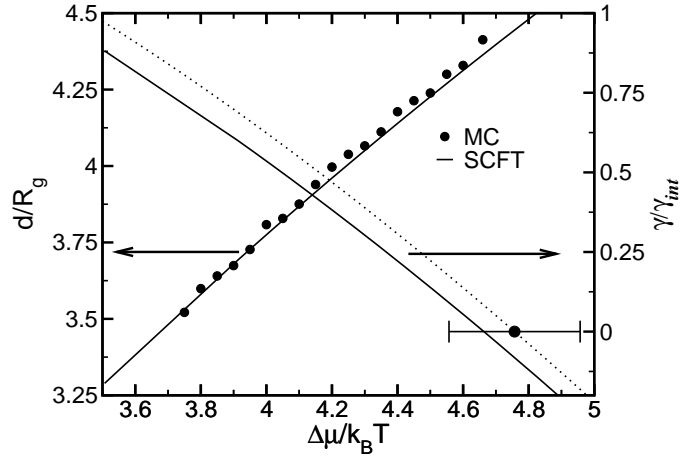}
\caption{Bilayer thickness, $d$, measured in units of the radius of 
gyration $R_g$ with scale to left, versus excess chemical potential as 
obtained in our SCFT calculation (solid line). This result is compared to 
those of an independent simulation averaged over all configurations, 
(solid circles). Also shown is the excess free energy per unit area, or 
tension, of the bilayer, scale to the right, as a function of the exchange 
chemical potential. Results obtained in our SCFT calculation are shown by 
the solid line, and those obtained in an independent simulation are shown 
by the dotted line. The error bar shows the uncertainty in the exchange 
chemical potential at which the bilayer is without tension as determined 
in the simulations. } \label{cpbilayer_SCFT} \end{figure}

The dependence of the membrane thickness, 
\begin{equation}
\label{thickness}
d=\delta \sigma_a V/(n_a+n_s),
\end{equation}
and membrane tension, $\gamma$, on the exchange chemical potential is
shown in Fig.~\ref{cpbilayer_SCFT}. The agreement is excellent between the
SCFT predictions and the independent simulation results averaged over all
configurations.  From the membrane tension, the area compressibility
modulus, $\kappa_A$, can be obtained by using any of the following
equivalent relations 
\begin{eqnarray}
\kappa_A&\equiv&A\frac{\partial\gamma}{\partial A}, \nonumber \\
       &= &-\delta\sigma_a\frac{\partial\gamma}{\partial\delta\sigma_a},
\nonumber \\
       &=&(\delta\sigma_a)^2\frac{\partial(\Delta\mu)}{\partial\delta\sigma_a}. 
\end{eqnarray}

Most of the earlier treatments of membranes relied on elasticity theory, 
in which a membrane is described solely by its elastic properties, such as 
the bending modulus, $\kappa_M$, the saddle-splay modulus, $\kappa_G$, and 
the spontaneous curvature, $c_0$ \cite{Safran94}. These moduli are 
normally taken either from an experimental measurement or from a 
microscopic theory. The SCFT approach, being based on a microscopic model, 
allows one to calculate these moduli in a straightforward manner which can 
be sketched as follows.

One calculates within the SCFT the excess free energies due to small 
spherical and cylindrical deformations of an interface containing an 
amphiphilic monolayer.  These excess free energies depend upon {\em 
microscopic} parameters, such as the amphiphile hydrophilic fraction $f$, 
as well as the curvatures of the deformations. They are then {\em fit} to 
the standard Helfrich Hamiltonian ${\cal H_E}$ of an infinitely thin 
elastic sheet, which depends upon {\em phenomenological} parameters, and 
the curvatures of the deformations. This Hamiltonian for a saturated, 
tensionless, monolayer is \cite {Safran94}

\begin{equation} {\cal H_E}=\int {\rm d}{\cal A}\; \left[ 2\kappa_M
(M-c_0)^2 + \kappa_G G \right], 
\end{equation}
where $M$ and $G$ are the local mean and Gaussian curvatures of the
deformed monolayer.  From the fit, one obtains the phenomenological
parameters in terms of the microscopic quantities. Details of this
procedure can be found in \cite{Matsen99} and \cite{Mueller002}. Because
SCFT ignores fluctuations, the moduli obtained are the bare,
unrenormalized values. The effect of renormalization is usually small
\cite{Peliti85} and depends on the lateral length scale of the
measurement. Fusion proceeds on a lateral length scale that does not
exceed the membrane thickness by a great deal and, on this length scale,
we expect the renormalization of the elastic constants to be small.

\begin{figure}[!ht]
\includegraphics[width=\fwidth]{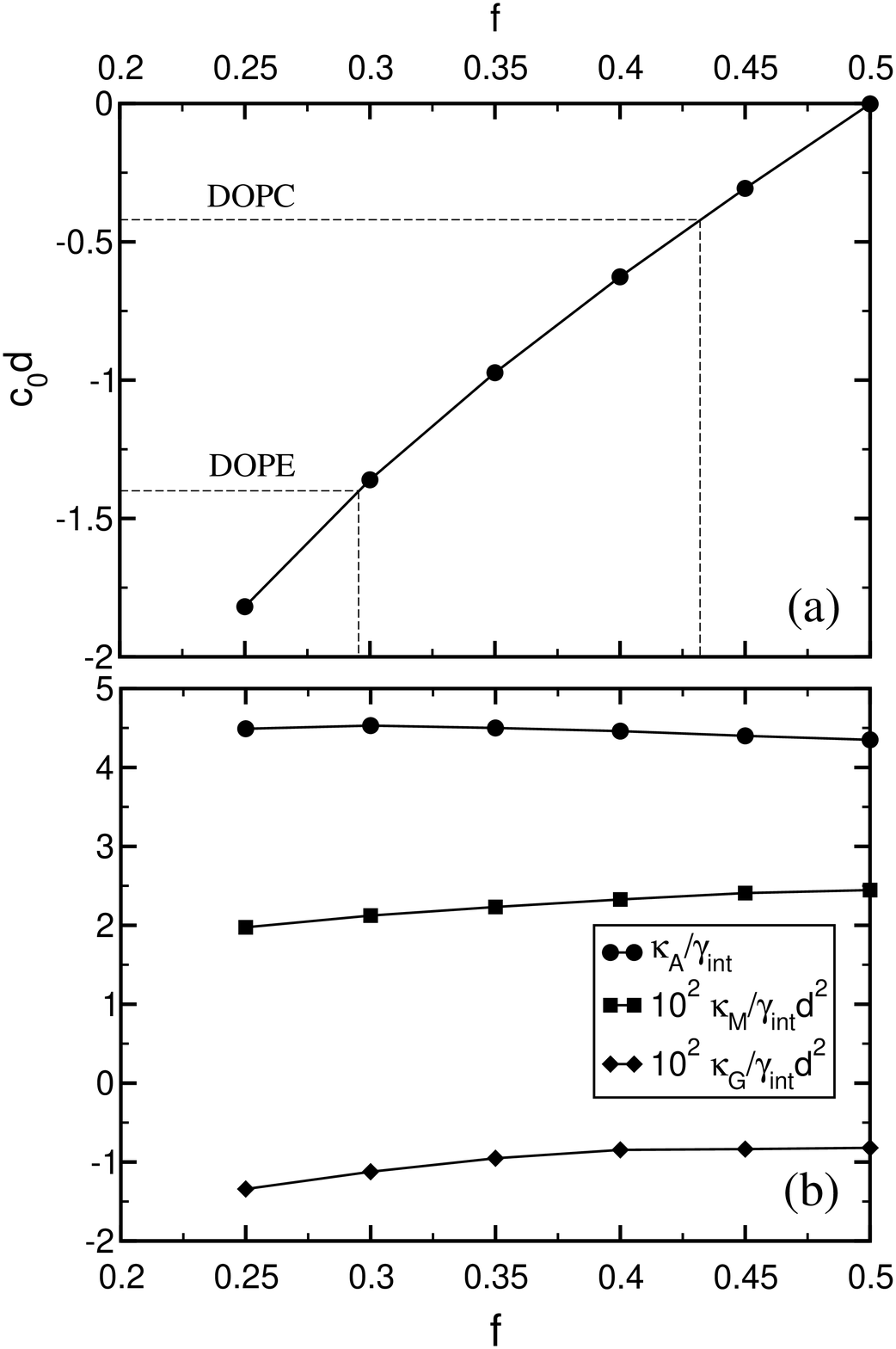}
\caption{(a) Dependence of the product of spontaneous curvature and 
bilayer thickness, $c_0d$, on hydrophilic fraction $f$ of the amphiphile. 
Experimental values of $c_0d$ for DOPE and DOPC from Table~\ref{table1} 
and the corresponding values of $f$ are shown. (b) Plot of the dependence 
of dimensionless values of the area compressibility, $\kappa_A$, bending 
modulus, $\kappa_M$, and saddle-splay modulus, $\kappa_G$, moduli on the 
hydrophilic fraction $f$ of the amphiphile.} \label{c0-vs-f-d} 
\end{figure}

In Fig.~\ref{c0-vs-f-d}(a) we show the calculated spontaneous curvature of 
an amphiphilic {\em monolayer} as a function of the hydrophilic fraction 
$f$ of the amphiphile.  
\highlight{The former is a monotonic function of the latter.  
This is to be expected as the phenomenological spontaneous curvature 
attempts to capture some of the effects of the differing hydrophilic and 
hydrophobic volumes in the amphiphile, which are specified by $f$, on the 
membrane configurations.} 
The above result provides a direct mapping 
between the phenomenological property $c_0$, which can be measured 
experimentally \cite{Rand90,Leikin96} and the microscopic variable $f$ 
used in our calculations.  Our results for the area compressibility, 
$\kappa_A$, and elastic moduli $\kappa_M$ and $\kappa_G$ are shown in 
Fig.~\ref{c0-vs-f-d}(b). In agreement with experiments on model lipid 
systems, they are not very sensitive to changes in amphiphile architecture 
parameter $f$ or, equivalently, to the spontaneous curvature $c_0$. For 
comparison with our model system, we present in Table~\ref{table1} the 
properties of lipid and amphiphilic diblock membranes determined 
experimentally.

Lipid membranes are known to exhibit strong mutual repulsion at small 
separations, usually attributed to so-called hydration forces 
\cite{Parsegian95}. We calculated the free energy of a system containing 
two planar bilayers in excess solvent under the condition that the 
distance between the {\em cis-} interfaces be constrained to a given 
value. This technique has been used before to study monolayer interactions 
in a similar system \cite{Thompson00}, and we refer the reader to this 
work for details.

One finds that there are two generic features of the free energy of the 
two bilayers as a function of their separation. First at sufficiently 
small separation, when heads of the amphiphiles come in contact, the 
membranes experience strong repulsion due to this contact, a repulsion 
which rises steeply as the separation is further reduced.  Second at a 
somewhat larger separation, there is a very weak attraction between the 
two membranes. These two features have been exhaustively analyzed 
\cite{Thompson00}. It has been determined that the repulsion arises mostly 
from the direct steric interaction between the head segments of the 
amphiphiles in the contacting {\em cis} monolayers. The weak attraction 
occurs because the solvent molecules prefer to leave the confining 
inter-membrane gap in order to increase their conformational entropy. This 
is simply the well-known depletion effect about which there is a very 
large literature \cite{Gotzelmann98}. The combination of the short-range 
repulsion and the depletion effect attraction produces a minimum in the 
free energy at some separation.  The distance at which this minimum 
occurs, typically on the order of one $R_g$ between opposing {\em cis} 
leaves, is the equilibrium separation.

The reasonable description of the properties of self-assembled monolayers 
and bilayers presented above provides confidence that our model can be 
used to describe the structural changes which occur in membranes during 
the fusion process.

\section{Energetics of the stalk and of its radial expansion} 
\label{sec:stalk-mech}

The formation and stability of the initial interconnection between the 
membranes, the stalk itself, has not received much theoretical attention. 
This is due to the fact that creating the stalk requires the membranes to 
undergo drastic topological changes, which can not be easily described by 
continuum elastic models. The common approach has been to assume that two 
small hydrophobic patches (one in each membrane) are produced by thermal 
fluctuations in the region of contact.  Hydrophobic interactions then 
drive the connection of these energetically costly regions, and result in 
the formation of a metastable structure, the stalk \cite{Markin83}.  The 
size of the hydrophobic patches and the distance between them can be 
optimized to minimize the free energy of the unstable transition state.  
With the use of this strategy, continuum elastic models have estimated the 
free energy barrier to create a stalk to be about $37k_BT$ 
\cite{Kuzmin01}. The SCFT method allows us to obtain solutions for {\em both}
the unstable transition state to stalk creation as well as for the stalk 
itself, if it is indeed metastable.

\begin{figure}[!ht]
\includegraphics[width=\fwidth]{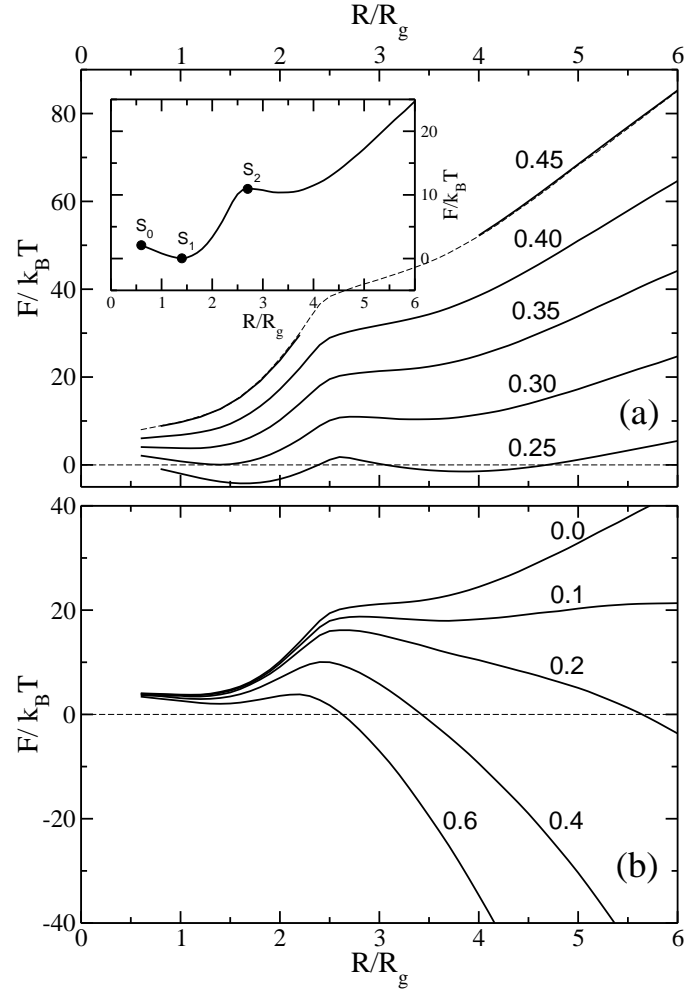}
\caption{(a) The free energy, $F$, of the stalk-like structure connecting 
bilayers of fixed tension, zero, is shown for several different values of 
the amphiphile's hydrophilic fraction $f$. In the inset we identify the 
metastable stalk, $S_1$, the transition state, $S_0$, between the system 
with no stalk at all and with this metastable stalk, and the transition 
state, $S_2$, between the metastable stalk and a hemifusion diaphragm. The 
architectural parameter is $f=0.30$ for this inset.  No stable stalk 
solutions were found for $f=0.45$ in the region shown with dashed lines. 
They were unstable to pore formation. (b) The free energy of the expanding 
stalk-like structure connecting bilayers of amphiphiles with fixed 
architectural parameter $f=0.35$ is shown for several different bilayer 
tensions. These tensions, $\gamma/\gamma_{\rm int}$, are shown next to 
each curve.} \label{stalk-zero-tension} \end{figure}

The dependence of the free energy of stalk-like structures on their 
radius, $R$, when the bilayers in which they form are under zero tension 
is shown in Fig.~\ref{stalk-zero-tension}(a). It is plotted for various 
values of the architectural parameter, from $f=0.45$, which corresponds to 
an amphiphile with a very small negative spontaneous curvature, to 
$f=0.25$, an amphiphile with a large negative spontaneous curvature. From 
Fig.~\ref{c0-vs-f-d}(a) it is seen that this $f$-range includes DOPC and 
DOPE lipids, which are frequently utilized as components of model lipid 
membranes in fusion experiments. At $f=0.45$ we could not find a 
stalk-like solution for $2.2<R/R_g<4$. The system would spontaneously 
rupture in the vicinity of the symmetry axis $r=0$, resulting in a fusion 
pore-like structure.

The extremal points of the free energy of the system with a stalk-like 
structure with respect to its radius correspond to intermediate structures 
and transition states.  The free energy extrema of the intermediates are 
saddle points with respect to deformations, and are therefore unstable, 
while those of the intermediates are local minima, and are therefore 
metastable. In the inset to Fig.~\ref{stalk-zero-tension}(a) we identify 
three states with extremal free energies.  Depending on the values of the 
architecture parameter $f$ and the membrane tension $\gamma$, we find the 
following solutions that play central roles in the description of the 
membrane fusion process:  the transition state $S_0$ between the 
unperturbed bilayers and the metastable stalk, the metastable stalk 
itself, $S_1$, and finally, a transition state which occurs as the initial 
small stalk is radially expanded into the hemifusion diaphragm. It is 
denoted $S_2$.

It is clear from Fig.~\ref{stalk-zero-tension}(a), that the metastable 
stalk solution $S_1$, and therefore the transition state $S_0$, exists 
only for sufficiently small values of the hydrophilic fraction $f$. At 
zero membrane tension this minimal $f$ is about $0.36$, which corresponds 
to the spontaneous curvature $c_0d=-0.9$, roughly that of a $1:1$ mixture 
of DOPE and DOPC lipids. To our knowledge, this is the first theoretical 
prediction of the metastability of the stalk itself. Existence of the 
stalk intermediate is crucial for the fusion dynamics, since it separates 
the fusion process into two activated stages: creation of the stalk and 
its further expansion. We argue, therefore, that for those conditions, 
specified by $f$ and $\gamma$, under which there is {\em no} $S_1$ 
solution, fusion will be considerably slower or, perhaps, impossible.  
Fig.~\ref{stalk-zero-tension}(a) also shows that, at small stalk radius, 
$R$, the free energy barrier to create the stalk, which corresponds to the 
state $S_0$, does not exceed $5k_BT$. This value is much smaller than the 
$37k_BT$ predicted by the phenomenological calculation \cite{Kuzmin01}. Of 
course, one can ask how applicable our results for block copolymer 
membranes are to biological ones. If we take as a natural measure of 
energy the dimensionless quantity $\gamma_{\rm int} d^2/k_BT$, where $d$ 
is the membrane thickness, we obtain a value of about $60$ in our system, 
a factor of 2.5 less than that ($\approx 150$) for a typical biological 
bilayer. This suggests that our energies should be scaled up by a factor 
of about 2.5 in order to compare them with those occurring in biological 
membranes. However even after doing so, they are still much smaller than 
the estimates of the phenomenological theories. We infer from this result 
that the phenomenological approaches based on the continuum elastic 
description are not accurate in describing the drastic changes in membrane 
conformations on the length scales comparable to the membrane thickness.

In Fig.~\ref{prof-stalk} (b), we show the calculated density profiles of 
the different segments in the unstable transition state structure $S_0$ at 
$f=0.35$ and $\gamma/\gamma_{\rm int}=0$. As can be seen, the interface 
between $A$ and $B$ segments is extremely curved and the stalk radius 
$R=0.6R_g$ is much smaller than the membrane hydrophobic core thickness. 
For clarity, we show only the majority component at each point. The 
interfaces between the different components appear to be sharp in such a 
graphical representation, but actually they are relatively diffuse, as can 
be seen in Fig.~\ref{comp-vs-z-LT}.

The behavior of the free energy of the stalk-like structure as it expands 
into a hemifusion diaphragm under {\em non-zero} tension is shown in 
Fig.~\ref{stalk-zero-tension}(b), and one clearly sees a second local 
maximum at $R/R_g\approx 3$ which corresponds to the second unstable 
transition state, $S_2$. The maximum results from the competition between 
the elimination of energetically-costly bilayer area, which reduces the 
free energy as $-\gamma \pi R^2$, and the creation of diaphragm 
circumference which increases it as $2\pi\lambda_{\rm tri} R$. Here 
$\lambda_{\rm tri}$ is a line tension. Fig.~\ref{stalk-zero-tension}(b) 
shows that this second maximum is, in general, much greater than that 
encountered in creating the stalk itself. Therefore we infer that crossing 
this barrier is the rate limiting step of the hemifusion mechanism.  As 
the tension increases, the height of this barrier decreases. Thus at very 
high tension, on the order of $0.5\gamma_{\rm int}$ for $f=0.3$, this 
local maximum disappears entirely leading to stalk expansion without any 
barrier at all. The rather strong dependence of this barrier height on 
tension contrasts with the behavior of the free energy of the metastable 
stalk itself, which, from the figure, is seen to depend only weakly on the 
tension. We also note that the minimum in the free energy corresponding to 
the metastable stalk is exceedingly shallow. This means that these 
interconnections are easily reversible, and would constantly fluctuate in 
size.

\begin{figure}[!ht]
\includegraphics[width=\fwidth]{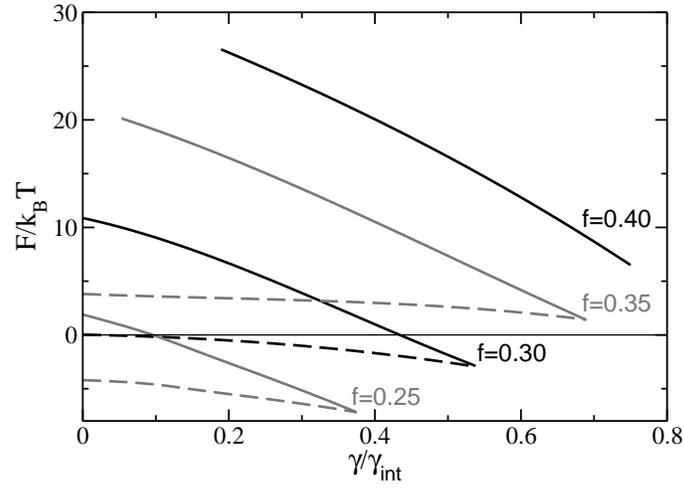}
\caption{ The free energy, $F$, of the metastable stalks $S_1$ (dashed 
lines) and the
  transition states $S_2$ (full lines)
  as a function of the tension for different
  architectures $f=0.25,\ 0.30,\ 0.35,\ 0.40$. Notice that there is no $S_1$
  solution for $f=0.4$ at the values of tension we studied 
}
\label{stalk-F-vs-tension}
\end{figure}

We recapitulate, in Fig.~\ref{stalk-F-vs-tension}, the dependence of the 
free energy of the metastable stalk $S_1$ and of the unstable transition 
state $S_2$ on the tension, $\gamma$, and on the architectural parameter 
$f$. The free energy of the metastable stalk ($S_1$) varies greatly with 
$f$, and decreases substantially for smaller $f$, i.e., as the spontaneous 
curvature of the amphiphile becomes more negative. Although our 
calculation applies to membranes composed of a single amphiphile, and not 
a mixture, it clearly strengthens the argument that one role of such 
negative curvature lipids as phosphatidylethanolamine, present in the 
plasma membrane, is to make metastable and thermally accessible the 
formation of stalks, which are necessary to begin the fusion process 
\cite{Dekruijff97} .  For sufficiently small values of the architectural 
parameter $f$, the free energy of the metastable stalk actually becomes 
lower than that of the unperturbed bilayers. Presumably this leads to the 
formation of the thermodynamically stable ``stalk phase'' recently 
realized experimentally in a lipid system \cite{Yang02}.

In summary, both an increase in the membrane tension $\gamma$ and a 
decrease in the hydrophilic fraction $f$ favor stalk expansion into the 
hemifusion diaphragm. The density profiles in the transition state between 
the metastable stalk and the hemifusion diaphragm is shown in 
Fig.~\ref{prof-stalk}(d), and that of the unstable, expanding, hemifusion 
diaphragm in Fig.~\ref{prof-stalk}(e). Note how thin is the hydrophobic 
region on the axis of symmetry in (d) and next to the triple junction in 
(e) compared to the thickness of the bilayers away from the diaphragm.

\section{Formation and expansion of fusion pore} \label{sec:pore} The 
hemifusion diaphragm is a possible intermediate along the path to fusion, 
but for complete fusion to occur, a hole must nucleate in the diaphragm 
leading to the formation of the final fusion pore. Here we consider the 
energetics of the fusion pore. The manner in which we obtain solutions 
corresponding to the pore and determine its free energy are similar to 
those for the stalk described above. A density profile of a fusion pore is 
shown in Fig.~\ref{prof-pore}(a).  In the plane of mirror symmetry between 
the two bilayers, there are two radii at which the volume fraction of 
hydrophobic and hydrophilic elements are equal, {\em i.e.} for which 
$\phi_A(r,0)-\phi_B(r,0)=0$.  We define the radius, $R$, of the fusion 
pore as the greater of these two distances.  This choice is consistent 
with the definition of the radius of a stalk-like structure, the only 
radius at which $\phi_A(r,0)-\phi_B(r,0)=0$, as can be seen from the 
following: if one begins with a stalk-like structure of radius $R$ and 
introduces a hole into it to create a pore, then the radius of the pore, 
according to the above definition, is also $R$, as it should be. We are 
able, therefore, to compare directly the free energies of the pore and 
stalk-like structures with the same radius.

\begin{figure}[!ht]
\includegraphics[width=\fwidth]{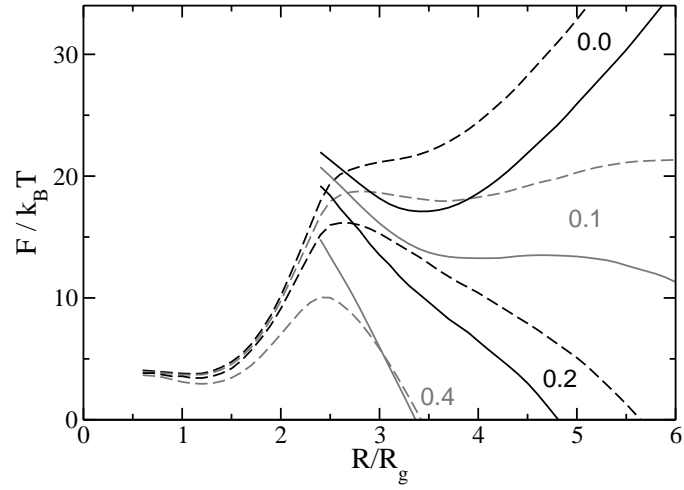}
\caption{The free energies, $F$, of a fusion pore (solid lines) and of a 
stalk (dashed lines, cf.~Fig.~\protect{\ref{stalk-zero-tension}}(b) of 
radius $R$ are shown.  Under the assumption that the stalk-like structure 
converts to a fusion pore when their free energies cross, one can read off 
the barrier between metastable stalk and formation of the fusion pore. The 
membranes are comprised of amphiphiles with fixed architecture, $f=0.35$, 
and are under various tensions. Note that the fusion barrier decreases 
with increasing tension. } 
\label{pore-F-vs-tension} \end{figure}

The free energy of the fusion pore in a system with fixed architecture, 
$f=0.35$, but under various tensions is shown in 
Fig.~\ref{pore-F-vs-tension}. As the tension on the membrane is increased 
from zero, the free energy of the pore decreases as $-2\gamma\pi R^2$ for 
large radius $R$, similar to the free energy decrease of the hemifusion 
diaphragm. The factor of two arises because pore expansion eliminates both 
membranes, whereas hemifusion expansion removes only one.

The free energies of the fusion pores, shown in solid lines, are compared 
with those of the stalk-like structures under the same conditions, which 
were shown previously in Fig.~\ref{stalk-zero-tension}(b).  We assume that 
a hole forms in the hemifusion diaphragm converting the diaphragm into a 
fusion pore when the free energy of the pore becomes lower than that of 
the stalk-like structure.  The radius at which the free energies of the 
expanding stalk and of the fusion pore become equal is $R\approx 
2.5-3.5R_g$. Of course, there is not a sharp transition from the stalk to 
the pore, because the free energies are finite. There is rather a region 
of radii where the transition from the stalk to the fusion pore occurs, 
and which is characterized by a free energy difference of the order of 
thermal fluctuations $k_BT$. Because the stalk and pore structures are so 
similar at such small values of radius, it is likely that there is only a 
very small free energy barrier associated with rupture of the hemifusion 
diaphragm, which converts it to the fusion pore. Therefore, our 
calculations indicate that the hemifusion diaphragm would hardly expand 
before the fusion pore would form.  This agrees with the conclusion of a 
recent phenomenological calculation \cite{Kuzmin01}.

Fig.~\ref{pore-F-vs-tension} shows that, except at low tension, the hole 
forms in the diaphragm {\em after} the barrier to diaphragm expansion has 
been crossed. Therefore within the standard hemifusion mechanism, the 
barrier to hemifusion expansion is, in fact, the major barrier to fusion.

\begin{figure}[!ht]
\includegraphics[width=\fwidth]{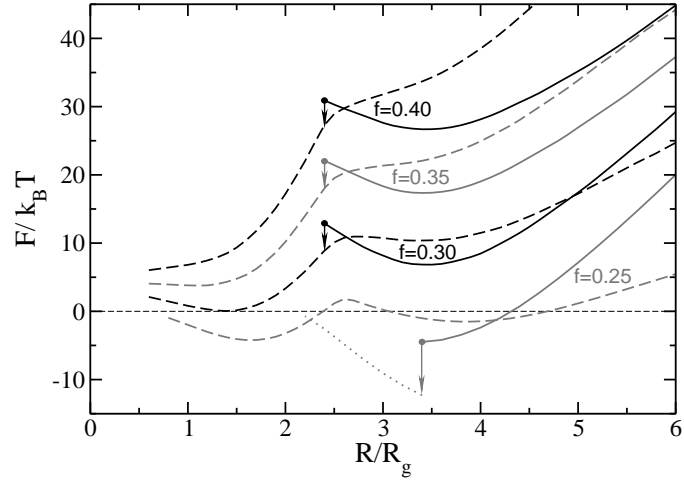}
\caption{The free energies, $F$, of a fusion pore (solid lines) and of a 
stalk
  (dashed lines) of radius $R$ are shown. In contrast to 
Fig.~\ref{pore-F-vs-tension}, the membranes here are under zero tension, 
and are comprised of amphiphiles with various values of $f$. The 
instability of the fusion pores at small radius is indicated by arrows. 
For $f=0.3$, 0.35, and 0.4, the stalk-like structure converts into a pore 
when it expands to a radius $R\approx 2.4R_g$ at which the free energies 
of stalk-like structure and pore are equal. For the system composed of 
amphiphiles of $f=0.25$, however, the stalk-like structure converts at 
$R\approx 2 R_g$ into an inverted micellar intermediate, (IMI), whose free 
energy is shown by the dotted line. The fusion pore is unstable to this 
IMI intermediate when its radius decreases to $R\approx 3.4R_g.$ Thus the 
IMI is the most stable structure under these conditions.}
\label{pore-zero-tension}
\end{figure}

At very low tensions, our results show that the most important barrier to 
fusion is no longer that governing the expansion of the hemifusion 
diaphragm, but becomes that associated with the expansion of the fusion 
pore itself. As a consequence of the large barrier to expansion of the 
fusion pore, pores of small radius, $R\approx 3.4R_g$, become {\em 
metastable} for most architectures for tensions $\gamma/\gamma_0$ less 
than about 0.1. This metastability persists to zero tensions, as seen from 
Fig.~\ref{pore-zero-tension}. Note that the potential minimum of the 
metastable pore is quite shallow, therefore one would expect that thermal 
fluctuations will easily cause it to expand and contract, or flicker, 
about the minimum free energy configuration with an amplitude of the order 
of $1R_g$. Note also that the activation barrier to reseal the fusion pore 
is quite substantial, on the order of $5k_BT$, which would translate to 
about $13k_BT$ in lipid systems. This means that flickering pores can be 
long-lived metastable states and, depending on the tension, might either 
reseal or expand. In retrospect, it should not be surprising that a small 
fusion pore is metastable in a tensionless membrane. Its free energy must 
increase linearly with its radius due to the line tension cost of its 
circumference. On the other hand as its radius decreases, the pore 
eventually becomes sufficiently small so that the inner sides of the pore 
come into contact, repelling each other and causing the free energy to 
again increase. Thus there must be a minimum at some intermediate 
distance. If the radius is decreased below some critical value, the 
compressed pore structure becomes unstable to the stalk-like structure. 
This instability is shown with arrows in Fig.~\ref{pore-zero-tension}. The 
density profile of such a marginally stable pore occurring in a membrane 
with $f=0.35$ and $R=2.4R_g$ is shown in Fig.~\ref{prof-pore}(a).

Fig.~\ref{pore-zero-tension} shows that the small flickering pore would be 
thermally excited from a metastable stalk for most of the architectures 
shown, $f=0.40$, 0.35, and 0.30. As the architecture changes such that $f$ 
decreases still further, the stalk free energy becomes negative meaning 
that it is favorable to create many of them which leads to the formation 
of a stalk phase. This occurs for non-zero tension as well.

Lastly we note from Fig.~\ref{pore-zero-tension} that at very low tension 
and very small values of the architectural parameter, $f=0.25$, 
corresponding to a large and negative spontaneous curvature, both the 
fusion pore and the stalk-like structure are unstable to another 
structure, the inverted micellar intermediate~(IMI), introduced by Siegel 
\cite{Siegel86}. Interestingly this configuration was suggested previously 
as a possible player in the fusion process, but has been largely neglected 
because free energy estimates obtained from elastic approaches were 
prohibitively high. As will be shown in our second paper, the IMI and 
related linear stalk intermediates are very important in describing 
fusion.

\section{Discussion} \label{sec:disc} 
\begin{figure}[!ht]
\includegraphics[width=\fwidth]{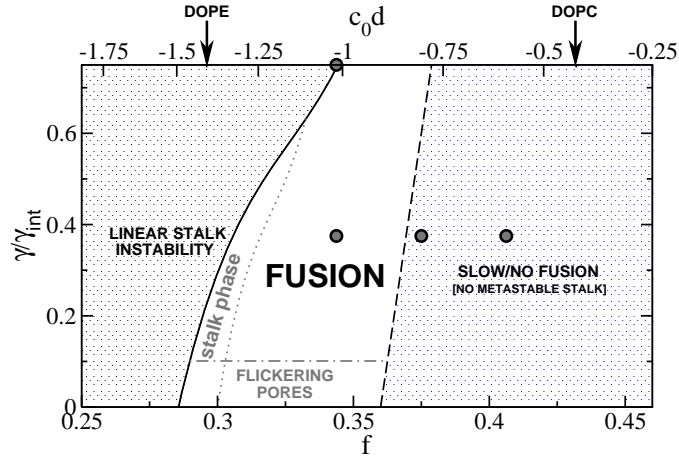}
\caption{A ``phase diagram'' of the hemifusion process in the hydrophilic 
fraction-tension, ($f,\gamma$), plane. Circles show points at which 
previous, independent, simulations were performed by us. Successful fusion 
can occur within the unshaded region. As the tension, $\gamma$, decreases 
to zero, the barrier to expansion of the pore increases without limit as 
does the time for fusion. As the right-hand boundary is approached, the 
stalk loses its metastability causing fusion to be extremely slow. As the 
left-hand boundary is approached, the boundaries to fusion are reduced, as 
is the time for fusion, but the process is eventually pre-empted due to 
the stability either of radial stalks, forming the stalk phase, or linear 
stalks, forming the inverted hexagonal phase. } 
\label{phases1} 
\end{figure}

We have carried out a 
self-consistent field study of the fusion of membranes consisting of 
flexible block copolymers in a solvent of homopolymer. The main purpose of 
this paper was to evaluate the free energy barriers encountered within the 
standard hemifusion, or stalk, mechanism.  We summarize our major findings 
in Fig.~\ref{phases1} in a form of a ``phase diagram'' in the parameter 
space of amphiphile architecture and tension.

We find that as the architecture of the amphiphile $f$ is changed so that 
its spontaneous curvature decreases from zero and becomes more negative, 
fusion is enhanced. There are at least two reasons for this. First, the 
initial stalk becomes {\em metastable} only if $f$ is sufficiently small, 
(i.e., the spontaneous curvature is sufficiently negative). This is the 
case in the region to the left of the nearly vertical black dashed line in 
Fig.~\ref{phases1}.  Presence of the metastable stalk intermediate is 
crucial for fast fusion, because its formation represents the first of at 
least two activated steps in the fusion process. Within the hemifusion 
mechanism, the second activated step is the radial expansion of the stalk 
into a hemifusion diaphragm. We expect that in the region in which the 
stalk is {\em not} metastable, fusion would be extremely slow, if not 
impossible.

We also find that for a sufficiently small $f$ the stalk intermediate 
becomes absolutely stable with respect to the system with unperturbed 
membranes. There is a small region (shown in the figure between the dotted 
and solid lines) in which the stalk is not only stable with respect to the 
unperturbed system, but is more stable than any other intermediate we have 
considered. Consequently we predict formation of a ``stalk phase'' in this 
region. This result is in accord with recent experiments on model lipid 
systems \cite{Yang02}.  As the spontaneous curvature is made even more 
negative, linear stalks, which are precursors to the formation of an 
inverted-hexagonal phase, become even more stable than radial stalks. 
Hence the stalk phase becomes unstable with respect to the inverted 
hexagonal phase.

We remark that, according to our calculations on bilayers comprised of a 
single amphiphile, {\em the variation of architectures within which 
successful fusion can occur is quite small}. One way that a system 
composed of many different amphiphiles can ensure the appropriate 
effective architecture is to employ a mixture of lipids comprised of those 
with small spontaneous curvatures, or lamellar formers, and those with 
larger negative spontaneous curvatures, or non lamellar-formers. Further, 
to remain within the small region of successful fusion, the ratio of these 
different kinds of lipids must be regulated, as indeed it is 
\cite{Morein96,Weislander97}.

We also note that the restrictions we have found on the possible variation 
of curvature, if fusion is to be successful, arise from stability 
arguments involving the stalk itself. Therefore any fusion mechanism which 
begins with the formation of a stalk will have this same phase diagram 
even though the subsequent path to fusion may differ radically from the 
standard hemifusion mechanism.

Within the range of the parameters where fusion can occur, we find that 
the activation barriers of the three major stages of the standard 
hemifusion process, which are stalk formation, stalk expansion, and pore 
formation, are affected differently by changes in architecture and in 
tension.  The first activation barrier is associated with creation of the 
initial metastable stalk. Its height is essentially independent of the 
membrane tension and increases only very weakly as $f$ is increased. Stalk 
creation is apparently {\em not} a rate limiting step, because the 
corresponding barrier does not exceed $5k_BT$ in our model, or $13k_BT$ if 
we extrapolate our results to lipid systems. This value is much lower than 
previous estimates of phenomenological theories \cite{Kuzmin01}.

In the second stage, during radial stalk expansion, the corresponding 
barrier is a very sensitive function of the amphiphile architecture, as 
can be seen in Fig.~\ref{stalk-F-vs-tension}. At small tension it ranges 
from $\approx 10k_BT$ for amphiphiles with small $f$ up to $\approx 
25k_BT$ for more symmetric amphiphiles. Again, this range translates into 
$25-63k_BT$ for biological bilayers. These observations are consistent 
with well known fusion enhancement effects of lipids with large and 
negative spontaneous curvature, such as DOPE, when they are added to 
fusing membranes \cite{Chernomordik95,Chernomordik96}. The effect of 
membrane tension on the stalk expansion barrier has also been determined. 
We find that increased tension lowers the activation barrier to radial 
stalk expansion and to pore formation.  This is in accord with experiment 
\cite{Monck90}. We predict that this barrier to fusion can, in principle, 
always be reduced and eliminated by sufficient tension. However, if the 
architecture of the amphiphiles is unfavorable, resulting in a very high 
barrier at zero tension, the tension needed to eliminate this barrier can 
be prohibitively high, and fusion will be pre-empted by membrane rupture.  
Results on the thermodynamics of membrane rupture will be presented in the 
second paper of this series.

The third stage of the process, nucleation of a hole in the diaphragm to 
convert the hemifusion diaphragm into a fusion pore and the pore's 
expansion, appears to present no additional barrier to fusion at most 
non-zero tensions. {\em Thus we find that the largest barrier encountered 
in this standard hemifusion mechanism is that associated with the 
expansion of the stalk into a hemifusion diaphragm}. The diaphragm does 
not expand very much, however, before this maximum barrier is reached.

At very low tensions, the controlling barrier is that associated with the 
expansion of the fusion pore. This leads to the prediction of the 
transient stability of small fusion pores. We believe that these 
structures correspond to ``flickering pores'' observed experimentally in 
lipid bilayer fusion \cite{Fernandez84,Spruce90,Chanturiya97}.

In spite of the agreement between our results on the hemifusion mechanism 
and the experimental observations mentioned above, there remain 
experimental observations which this hypothesis does not explain. First, 
lipid mixing is observed to occur not only between the {\em cis} 
monolayers, but also between {\em cis} and {\em trans} layers 
\cite{Lentz97}. Second, transient leakage is observed, and it is 
correlated spatially and temporally with fusion \cite{Frolov03}. In 
addition to these experimental observations, simulation studies have 
revealed, both in planar bilayer fusion and in vesicular fusion, a very 
different pathway subsequent to the formation of the initial metastable 
stalk. To understand these discrepancies, we have performed calculations 
similar to the ones presented here for the alternative mechanism proposed 
recently \cite{Noguchi01,Mueller02,Mueller03}. These results will be 
presented in the second paper of this series.

\begin{acknowledgments} 
\vspace{-.8\baselineskip} 
We acknowledge very useful conversations with L. Chernomordik, F. Cohen,
M. Kozlov, B. Lentz, D. Siegel, and J. Zimmerberg. We are particularly
grateful to V. Frolov for sharing his knowledge and expertise with us.
Financial support was provided by the National Science Foundation under
grant DMR 0140500 and the DFG Bi314/17 and Mu1674/1 . Computer time at the
NIC J{\"u}lich, the HLR Stuttgart and the computing center in Mainz are
also gratefully acknowledged \end{acknowledgments}

\section{Appendix}
\subsection{Self-consistent field theory for polymer systems}
\label{sec:SCFT-model}
The first step in any field-theoretic approach to such  systems is to
convert the partition sum over all possible molecular configurations into
an integration over configurations of corresponding, smooth, collective
variables, the density and chemical potential fields.  The
derivation of the effective field-theoretic Hamiltonian of our polymer
model follows a standard prescription, so we give only the final
expression here:
\begin{eqnarray}
\frac{{\cal H}[T,\Delta \mu, V, {\cal
A};w_A,w_B,\phi_A,\phi_B,\xi]}{k_BT\Phi}
= 
&-& Q_s[w_A] - z Q_a[w_A,w_B]\nonumber\\
&+& \chi N\int {\rm d}V\; \phi_A(\bfr)\phi_B(\bfr)\nonumber\\
&+& \int {\rm d}V\; (\phi_A(\bfr)w_A(\bfr) + \phi_B(\bfr)w_B(\bfr))\nonumber\\
&+& \int {\rm d}V\; \xi(\bfr) (\phi_A(\bfr) + \phi_B(\bfr) - 1).
\label{FTH}
\end{eqnarray} 
Here $z\equiv \exp{\Delta\mu/k_BT}$ is the {\em relative} activity of the 
amphiphiles, where $\Delta\mu=\mu_a -\mu_s$, the difference in bulk 
chemical potentials of the amphiphile and the solvent. There is only one 
independent chemical potential because the liquid is assumed to be 
incompressible. The fixed number density of polymer chains, $(n_a+n_s)/V$ 
is denoted $\Phi$.  Thus the total number density of chains is fixed, but 
the relative amounts of amphiphile and solvent chains is controlled by the 
chemical potential difference $\Delta\mu$. The Flory interaction 
parameter, $\chi$, is inversely proportional to the temperature. The 
$Q_s[w_A]$ and $Q_a[w_A,w_B]$ are the single chain partition functions of 
the solvent and amphiphile molecules subjected to the local chemical 
potential fields $w_A(\bfr)$ and $w_B(\bfr)$, which act on the $A$ and $B$ 
segments respectively. The local volume fractions of $A$ and $B$ monomers 
are given by $\phi_A(\bfr)$ and $\phi_B(\bfr)$. The ``local pressure'' 
$\xi(\bfr)$ is the Lagrange multiplier field introduced to enforce the 
incompressibility condition.

The mean field approximation, which is at the heart of the SCFT approach, 
amounts to extremizing the field-theoretic Hamiltonian with respect to all 
the fields upon which it depends. Fluctuations about this extremum are 
ignored. In systems of relatively long chains, in which composition 
fluctuations are small, the results of this approximation are very good 
indeed \cite{Bates99}.

It can be shown that the field configurations that correspond to
stationary points of ${\cal
H}[T,\Delta\mu,V,{\cal A};w_A(\bfr),w_B(\bfr),\phi_A(\bfr),
\phi_B(\bfr),\xi(\bfr)]$, denoted in
the following by an over-bar, satisfy the following set of coupled
non-linear equations:

\begin{eqnarray}
\bw_A(\bfr) &=& \chi N \bphi_B(\bfr) + \bxi(\bfr)\label{wa}\\
\bw_B(\bfr) &=& \chi N \bphi_A(\bfr) + \bxi(\bfr)\label{wb}\\
1 &=& \bphi_A(\bfr) + \bphi_B(\bfr)\label{incompress}\\
\bphi_A(\bfr) &=& \int_0^1 {\rm d}s\; q_s(\bfr,s)q_s(\bfr,1-s)
+ z\int_0^f {\rm d}s\; q_c(\bfr,s)q_c^\dag(\bfr,s)\label{adensity}\\
\bphi_B(\bfr) &=& z\int_f^1 {\rm d}s\; q_c(\bfr,s)q_c^\dag(\bfr,s),\label{SCFT-eqn}
\end{eqnarray}
where the single chain propagators $q_c(\bfr,s)$,$q_c^\dag(\bfr,s)$, 
and $q_s(\bfr,s)$ satisfy the usual modified diffusion equations for the
flexible polymer chains with the Gaussian statistics; e.g. for the homopolymer
solvent it is
\begin{equation}
\frac{\partial q_s(s)}{\partial s}=
R_g^2\nabla^2 q_s(\bfr,s)-\bw_A(\bfr)q_s(\bfr,s), \,\,\,\mbox{with}\,\,
q_s(\bfr,s=0)=1.
\label{prop-eqn}
\end{equation}
Here $R_g$ is the radius of gyration of the unperturbed Gaussian polymer.
The value of the free energy of the stationary configuration is given
simply by 
\begin{equation} 
\Omega_m(T,\Delta\mu,V,{\cal A}) = 
{\cal H}[T,\Delta\mu,V,{\cal A};\bw_A,\bw_B,\bphi_A,\bphi_B,\bxi]
\end{equation}

The system of non-linear equations~(\ref{wa})-(\ref{SCFT-eqn}) together 
with the equations for the propagators~(\ref{prop-eqn}) can be solved 
numerically in real space by a straightforward relaxational iterative 
algorithm \cite{Drolet99}. The real-space approach has far more 
flexibility in studying localized structures, such as the fusion 
intermediates, than does the more numerically optimized spectral approach 
\cite{Matsen94}.  The recently proposed pseudo-spectral techniques 
\cite{Rasmussen02} appear to be very efficient, but they rely heavily on 
the numerical fast Fourier transforms, which are not available for the 
cylindrical coordinates, ${\bf r}=(r,\varphi,z)$, we use in our 
calculations. Because the local volume fractions are required to be 
axially symmetric, $\phi_A(r,z)$, $\phi_B(r,z)$, and to be symmetric under 
reflection in the $z=0$ plane, the problem is two-dimensional and need 
only be solved in one quadrant. We impose reflecting boundary conditions 
at $z=0$, $z=z_{\rm max}$, $r=0$ and $r=r_{\rm max}$, where $z_{\rm max}$ 
and $r_{\rm max}$ determine the size of the computational cell. They were 
set to $8R_g$ and $15R_g$ respectively. We discretized all the fields on a 
uniform lattice with resolution $\Delta r=\Delta z=0.1-0.05R_g$. In 
solving the diffusion equation (\ref{prop-eqn}) we used contour length 
discretization of $\Delta s=0.01-0.001.$ This gave us an accuracy of no 
less than $0.1k_BT$ in the free energy. On a one GHz, Pentium 3, work 
station, approximately five minutes were required to obtain convergence 
for a configuration, $\phi_A(r,z)$ and $\phi_B(r,z)$, at a given value of 
$f$ and $\gamma$.
 
To describe a stalk-like structure, the solutions must smoothly connect 
the hydrophobic regions of the two bilayers.  The radius, $R$, of the 
stalk, again defined by $\phi_A(R,0)-\phi_B(R,0)=0$, must take a value 
that we specify. To facilitate finding such a solution we apply, at the 
beginning of our calculations, auxiliary external fields which favor 
hydrophobic segments along the axis of symmetry between the two {\em 
trans} leaves.  After the system assembles into the desired structure, 
these auxiliary external fields are switched off, and the solution is 
allowed to relax.  Far from the axis of symmetry, the membranes reach 
their equilibrium separation. Again this minimum is due to the short-range 
repulsion and the attractive depletion interaction between the bilayers. 
Once a solution satisfying these constraints is obtained, the free energy 
of the stalk-like structure is then calculated. It is the difference, at 
constant chemical potential (or, equivalently, constant tension), between 
the free energy of the system with two bilayers connected by the 
stalk-like structure, and that of the two unperturbed bilayers without the 
interconnection.

\subsection{Reaction coordinate constraint}
\label{sec:constraint}
The SCFT strategy and its numerical implementation along the lines 
presented above are capable, in practice, only of identifying 
thermodynamic, locally stable, configurations of the system. To clarify 
this point, consider the following construction. Suppose, we first 
extremize ${\cal H}$ with respect to the chemical potential fields 
$w_A(\bfr)$ and $w_B(\bfr)$, and the incompressibility field $\xi(\bfr)$. 
Then, at least in principle, we obtain a free energy functional that 
depends only on the physical density fields $\phi_A(\bfr)$ and 
$\phi_B(\bfr)$: \begin{equation} {\cal F}[\phi_A,\phi_B]\equiv 
\mbox{extremum}_{\{w_A,w_B,\xi\}} {{\cal H}[w_A,w_B,\phi_A,\phi_B,\xi]}, 
\end{equation} where we have suppressed the dependence on the 
thermodynamic variables, $T,\ \Delta\mu,\ V$ and ${\cal A}$. An extremum 
of ${\cal F}[\phi_A,\phi_B]$ is the system's free energy $\Omega_m$. This 
extremum corresponds to a thermodynamic, locally stable, state if and only 
if the matrix of the second derivatives $\delta^2{\cal 
F}/{\delta\phi_A(\bfr)\delta\phi_B(\bfr')}$ is positive definite in that 
configuration, i.e., the locally stable configurations correspond to the 
{\em minima} of the free energy density functional ${\cal 
F}[\phi_A,\phi_B]$. To distinguish them from other kinds of solutions, we 
will refer to these locally stable structures as {\em intermediates}.

In studying an activated process, such as membrane fusion, we need to know 
not only the free energy of the intermediates along some reaction path, 
but also the properties of the {\em transition states}, which correspond 
to {\em saddle} points of the free energy functional ${\cal 
F}[\phi_A,\phi_B]$. Unfortunately, finding a saddle point of a functional 
poses a serious numerical problem. In particular, commonly used 
relaxational algorithms prove to be inadequate for this task, because they 
rely on local stability around a solution, which is obviously lacking at 
the saddle point. Newton-Raphson type algorithms are capable of finding 
any extremal points as long as the initial configuration is within the 
basin of attraction of that point. Unfortunately, we do not know the 
location of a saddle point in advance, so these methods are also not very 
practical.

In some cases one can identify the unstable directions of the functional
and stabilize it by applying a suitable constraint.  As an example,
consider a transition state between the stalk and the hemifusion
diaphragm. We treat this situation in detail in the main text and use it
here simply for illustration.  We confine the possible solutions to be
axially symmetric and to have a mirror symmetry with respect to $z=0$
plane because the unstable mode of either the stalk or diaphragm can be
associated with an overall radial contraction or expansion of the
structure. We further constrain the solution by requiring that, in the
$z=0$ plane, the $A/B$ interface (i.e., the locus of points at which
$\phi_A(\bfr)-\phi_B(\bfr)=0$) be located on a circle of some specified
radius $R$. This radius plays the role of a reaction coordinate. To impose
the constraint, we employ a Lagrange multiplier, $\psi_R$ \cite{Matsen99}.
The corresponding constrained field-theoretic Hamiltonian is then given by
\begin{equation} \frac{{\cal H}_R[w_A,w_B,\phi_A,\phi_B,\xi,
\psi_R]}{k_BT\Phi}\equiv
 \frac{{\cal H}[w_A,w_B,\phi_A,\phi_B,\xi]}{k_BT\Phi}
-\psi_R \int {\rm d}V\; \delta(z)\delta(r-R)(\phi_B(\bfr)-\phi_A(\bfr)).
\label{FTHc}
\end{equation}
 The first two SCFT Eqns.~(\ref{wa}) and (\ref{wb}) 
should be modified accordingly: 
\begin{eqnarray}
w_A(\bfr) &=& \chi N \phi_B(\bfr) + \xi(\bfr) + \psi_R\delta(z)\delta(|\bfr|-R)
\nonumber\\
w_B(\bfr) &=& \chi N \phi_A(\bfr) + \xi(\bfr) - \psi_R\delta(z)\delta(|\bfr|-R)
\label{w-constraint}
\end{eqnarray}
and the third equation expressing the local density constraint is
supplemented by the additional local constraint:
\begin{equation}
0 = \left(\phi_B(r, z)-\phi_A(r, z)\right)|_{r=R,z=0},
\end{equation}
where, again,  $(r,z)$ are cylindrical coordinates. The Lagrange multiplier
$\psi_R$ plays a role of the local chemical potential that couples to the
density fields. The solution of the SCFT equations optimizes the free
energy with respect to $\psi_R$ as well as the other fields. The same
relaxational iterative approach as that used for the non-constrained case
proved to be efficient for solving this modified problem.

In general, for an arbitrary position of the constraint $R$, the value of 
the Lagrange multiplier $\psi_R$ is non-zero, which means that the 
corresponding field configuration is not a solution of the original 
non-constrained system.  Nevertheless, at the points where $\psi_R$ {\em 
is} zero, the constrained and non-constrained sets of equations become 
identical, i.e., an extremum of the constrained free energy functional is 
also an extremum of the non-constrained functional.

The free energy of the constrained system, $F_R$, quite generally,
satisfies the following ``force balance'' equation:
\begin{equation}
\frac{{\rm d}F_R}{{\rm d}R} = -2\pi R\psi_R
\left.\frac{{\rm d}[\phi_B(r,z)-\phi_A(r,z)]}{{\rm d}r}\right|_{r=R,z=0}
\end{equation}
The condition $\psi_R=0$ is equivalent to $dF_R/dR=0$, so that the 
extremal points of $F_R$ as a function of $R$ are also the extremal points 
of ${\cal F}[\phi_A,\phi_B]$ as a function of $R$. Clearly, a minimum of 
free energy $F_R$ of the restricted system corresponds to a minimum of 
${\cal F}[\phi_A,\phi_B]$, whereas a maximum of $F_R$ corresponds to a 
saddle-point configuration of ${\cal F}[\phi_A,\phi_B]$. By scanning a 
range of $R$ values, we can identify all the metastable intermediates and 
unstable transition states along a particular path. Moreover even those 
constrained configurations for which $\psi_R\neq 0$ have a transparent 
physical interpretation, and provide a wealth of additional information, 
inaccessible by the non-constrained calculations.  For other applications 
of similar constraints we refer the reader to the literature 
\cite{Matsen99,Mueller0002,Duque03}.

\subsection{Model parameters} To make a direct comparison with our 
previous independent Monte Carlo simulations \cite{Mueller03}, we match 
corresponding model parameters. The length scale in SCFT calculations is 
usually set by the polymer radius of gyration $R_g$. For the polymers in 
the simulations, the radius of gyration was found to be $R_g=6.93u$, where 
$u$ is the spacing of the cubic lattice into which the simulation volume 
is divided. Because of the incompressibility constraint, the volume per 
polymer, $V/(n_a+n_s)$, with $n_a$ and $n_s$ the number of amphiphilic and 
solvent polymers respectively, enters the SCFT free energy only as a 
multiplicative factor and in a dimensionless ratio $V/[(n_a+n_s)R_g^3]$. 
As this ratio was taken to be $1.54$ in the simulations, we take the same 
value in evaluating the SCFT free energy. The energy scale in SCFT is set 
by the product of the Flory interaction parameter $\chi$ and the 
polymerization index $N$.  It has been shown previously that the choice 
$\chi N=30$ corresponds to the simulated system \cite{Mueller03}.

\clearpage
\bibliography{kkfusion03}

\clearpage

\begin{table}
\begin{tabular}{cccc}
\hline \hline
          & Polymersomes       & \multicolumn{2}{c}{Liposomes}      \\  
          &  EO7               & DOPE              &     DOPC\\   \hline
$d$     & 130\AA               & $38.3\AA^{(a)}$   & $35.9\AA^{(b)}$ \\
$c_0 d$ & no data              & $-1.4^{(d)}$      & $-0.42^{(c)}$ \\
$\kappa_A/\gamma_{\rm int}$
          & 2.4                & $4.4^{(b)}$       & $2.9^{(b)}$    \\
$\kappa_M/\gamma_{\rm int} d^2\times 10^2$
          & 1.67               & $6.0^{(c)}$       & $6.0^{(d)}$   \\
\hline \hline
\end{tabular}
\caption{Structural and elastic properties of bilayer membranes:
$d$ - bilayer membrane thickness,
$c_0$ - monolayer spontaneous curvature,
$\kappa_A$ - bilayer area compressibility modulus,
$\kappa_M$ - monolayer bending modulus,
$\gamma_{\rm int}=50pN/nm$ - oil/water interfacial tension.
Data on EO7 polymersomes is taken from \onlinecite{Discher99}; and on lipids
from (a): \onlinecite{Rand89},  (b): \onlinecite{Rand90},
(c): \onlinecite{Chen97}, and (d): \onlinecite{Leikin96}
(see also http://aqueous.labs.brocku.ca/lipid/).
Values of $d$, $c_0$ and $\kappa_a$ for DOPE were obtained by linear
extrapolation from the results on DOPE/DOPC(3:1) mixtures and pure DOPC.
}
\label{table1}
\end{table}

\clearpage

\end{document}